\documentclass[12pt]{article}
\usepackage{epsf}
\usepackage{graphicx}
\usepackage{amssymb}
\usepackage{bbm}

\def\d{{\mathrm{d}}}
\def\Tr{{\mathrm{Tr}}}
\def\id{\mathbbm{1}}

\def\CC{{\mathcal C}}
\def\CD{{\mathcal D}}

\def\CF{{\mathcal F}}
\def\CG{{\mathcal G}}
\def\CH{{\mathcal H}}
\def\CK{{\mathcal K}}

\def\CP{{\mathcal P}}
\def\CS{{\mathcal S}}

\def\MC{{\mathbf C}}
\def\MG{{\mathbf G}}
\def\MH{{\mathbf H}}
\def\MK{{\mathbf K}}
\def\MW{{\mathbf W}} 
\def\MZ{{\mathbf Z}}
\def\ag{{\mathrm{ag}}}
\def\cg{{\mathrm{cg}}}
\def\diag{{\mathrm{diag}}}
\def\mag{{\mathrm{mag}}}

\date{\today}

\def\rene{\renewcommand{\arraystretch}{1.8}}

\rene

 \newcommand{\mysection}[1]{\section{#1}
                    \setcounter{table}{0}\setcounter{equation}{0}}

\begin{document}

\begin{titlepage}
\begin{flushright}
\normalsize UNITU--THEP--17/2000
\end{flushright}

\vspace{2cm}

\par
\vskip .5 truecm
\large 
\begin{center}
{\bf Abelian and center gauge fixing \\
in continuum 
Yang-Mills-Theory for general gauge groups\footnote{supported by DFG 
under grant-No. DFG-Re 856/4-1 and DFG-EN 415/1-2}}
\end{center} 
\par
\vskip 1 truecm
\normalsize
\begin{center}
{\bf H.~Reinhardt}\footnote{e--mail: \tt  reinhardt@uni-tuebingen.de}
$^{{\mathrm{a,b)}}}$ and 
{\bf T.~Tok}\footnote{e--mail: \tt tok@alpha6.tphys.physik.uni-tuebingen.de}
$^{{\mathrm{a)}}}$
\\
\vskip 1 truecm
\it{a) Institut f\"ur Theoretische Physik, Universit\"at T\"ubingen\\
Auf der Morgenstelle 14, D-72076 T\"ubingen, Germany}
\\
\vskip 0.5 truecm
\it{b) Center for Theoretical Physics\\
Laboratory for Nuclear Science and Department of Physics\\
Massachusetts Institute of Technology\\
Cambridge, Massachusetts 02139}
\\
\today

\end{center}
 \par
\vskip 2 truecm\normalsize
\begin{abstract} 
A prescription for center gauge fixing for pure Yang-Mills theory
in the continuum with general gauge groups is presented.
The emergence of various types of singularities (magnetic monopoles and center
vortices) appearing in the course of   the gauge fixing procedure 
are discussed.
\end{abstract}



\end{titlepage}
\baselineskip=20pt

\mysection{Introduction}

There are two promising mechanisms to explain color confinement: 
the dual Meissner effect \cite{Parisi,Mandelstam,tHooft-76-1} 
and the picture of condensation of center vortices 
\cite{tHooft-78,Mack-79}. 
The relevant infrared degrees of freedom corresponding to these 
two mechanisms are magnetic monopoles and center vortices, respectively. 
Both can be identified in partial gauge fixings where they arise as
defects to the gauge fixing, magnetic monopoles and center vortices 
respectively arise in Abelian gauges and center gauges. 
Both pictures of confinement have received support from recent 
lattice calculations performed in specific gauges to identify the relevant 
infrared degrees of freedom.   
Monopole dominance in the string tension \cite{Suzuki-90,Hioki-91,Bali-98} 
has been found in maximally Abelian gauge and in all forms of the
Abelian gauges considered monopole condensation occurs in the 
confinement phase and is absent in the de-confinement phase 
\cite{DiGiacomo-00}. 
Lattice calculations performed in the so-called maximum center
gauge show that the vortex content detected after center projection 
produces virtually the full string tension, while the string tension 
disappears, if the center vortices are removed from the lattice ensemble 
\cite{DelDebbio-97,Forcrand-99-1}. This property of center dominance 
exists at finite temperature \cite{Langfeld-99}.
The vortices have also been shown to condense in the 
confinement phase \cite{Kovacs-00}. Furthermore in  the gauge field 
ensemble devoid of center vortices chiral symmetry breaking disappears 
and all field configurations belong to the topologically trivial sector
\cite{Forcrand-99-1}. The continuum version of maximum center gauge 
has been derived in \cite{Engelhardt-00-1}.

Both gauge fixing procedures, the maximally Abelian gauge as well as the
maximum center gauge, suffer from the Gribov problem \cite{Gribov-78}.
To circumvent the Gribov problem, the Laplacian gauge \cite{Vink-92},
the Laplacian Abelian gauge \cite{vdSijs-96}  
and the Laplacian center gauge 
\cite{Alexandrou-99-1,Alexandrou-99-2}  
have been introduced. 
In the Laplacian Abelian and Laplacian center gauge one uses
(usually the two lowest-lying) eigenfunctions $\psi_1$ and $\psi_2$ 
of the covariant Laplace operator 
transforming in the adjoint representation of the gauge group to 
define the gauge. The Laplacian center gauge can be understood as the 
extension of the Laplacian Abelian gauge. For gauge group $SU(2)$ the 
Abelian gauge is fixed by demanding 
that at every point $x$ in space-time the gauge fixed field $\psi_1^V(x)$ 
points into the positive $3$-direction in color space. To fix the residual
Abelian gauge freedom up to the center ${\mathbbm{Z}}_2$ of the gauge 
group one 
rotates the second to lowest eigenvector $\psi_2^V(x)$ 
into the $1-3$-plane in color space. Actually it is not important that $\psi_1$
and $\psi_2$ are eigenfunctions of the covariant Laplace operator -
the only important point is that $\psi_1$ and $\psi_2$ 
homogeneously transform under gauge transformations. 
Merons and instantons configurations have been studied in this gauge in
\cite{Reinhardt-00}. Recently de~Forcrand and Pepe \cite{deForcrand-00-1} 
extended the Laplacian center gauge 
to $SU(N)$ gauge groups. The aim of the present paper is to further
extend the Laplacian center gauge to arbitrary gauge groups. 
As an example we consider the symplectic group 
$Sp(2)$ which is the universal covering group of $SO(5)$.
The latter group is relevant in connection with string theories and
superconductivity \cite{SO(5)}

\mysection{Lie algebra conventions}

We denote by $\MG$ and $\CG$ the gauge group and its Lie algebra,
respectively, and by $\MH \subset \MG$ and $\CH \subset \CG$ 
its Cartan subgroup and subalgebra, respectively. The group $\MG$ 
is assumed to be simple and has rank $r$. We use the following 
Lie algebra conventions. We denote by 
$H_k \, , \, k = 1 , \ldots , r$ an orthogonal basis in 
the Cartan subalgebra $\CH \subset \CG$ normalized to
\begin{equation}
\label{norm}
\Tr ( H_k H_l ) = \delta_{k l} \, ,
\end{equation} 
and satisfying
\begin{equation}
\label{commutation-1}
\left[ H_k , E_\alpha \right] = \alpha_k E_\alpha \, ,
\end{equation}
where $E_\alpha $ is the root vector\footnote{We work in the
complexified Lie algebra. The real Lie Algebra is spanned by the
elements $i H_k$, $i ( E_\alpha + E_{-\alpha})$ and 
$(E_\alpha - E_{-\alpha})$.} to the root $\alpha$.
We denote by Greek letters $\rho, \sigma , \ldots $ vectors 
in the Cartan subalgebra
\begin{equation}
\rho = \sum_{k=1}^r \rho_k H_k \in \CH \, , \quad 
\rho_k \in {\mathbbm{R}} \, , \quad k = 1 , \ldots , r 
\end{equation}
for which we define the scalar product by
\begin{equation}
( \rho , \sigma ) = \Tr ( \rho \sigma ) = 
\sum_{k=1}^r \rho_k \sigma_k
\end{equation}
where the last relation follows from (\ref{norm}). Furthermore, let
\begin{equation}
\alpha_{(i)} \, , \quad 
\mu_{(i)}  \, , \quad 
\alpha^{\vee}_{(i)} = 
\frac{2 \alpha_{(i)}}{(\alpha_{(i)},\alpha_{(i)})}\, , \quad 
\mu^{\vee}_{(i)} = 
 \frac{2 \mu_{(i)}}{(\alpha_{(i)},\alpha_{(i)})}
\end{equation}
be the simple roots, fundamental weights, co-roots and co-weights,
respectively, satisfying
\begin{equation}
\label{roots-weights-2}
( \alpha^{\vee}_{(i)} , \mu_{(j)} ) = 
(\alpha_{(i)} , \mu^{\vee}_{(j)} ) = 
\delta_{i j} \, .
\end{equation}
We denote by $\Sigma$, $\Sigma^+$ and $\Pi$ the sets of all
roots, all positive roots and all simple roots, respectively. 
For later use we also note that the co-weight vectors 
$\mu^{\vee}_{(i)}$ generate the center elements via
\begin{equation}
\label{center-1}
z_i = \exp \left( 2 \pi i \mu^{\vee}_{(i)} \right) \, ,
i = 1 , \ldots , r
\end{equation}
which form the center of of the group $\MG$, whereas for the 
co-roots $\alpha^{\vee}_{(i)}$ we have
\begin{equation}
\label{center-2}
\id = \exp \left( 2 \pi i \alpha^{\vee}_{(i)} \right) \, ,
i = 1 , \ldots , r \, .
\end{equation}
Note that the lattice generated by the co-roots 
$ \alpha^{\vee}_{(i)} \, , \, i = 1 , \ldots , r $ is a subset of 
the lattice generated by the co-weights 
$ \mu^{\vee}_{(i)} \, , \, i = 1 , \ldots , r $.

\mysection{Abelian gauge fixing}

Before presenting the Laplacian center gauge fixing for arbitrary Lie
groups it is worth while to introduce its Abelian counter part, 
the (Laplacian) Abelian gauge fixing for general Lie groups, which is
part of the (Laplacian) center gauge fixing.
To this end we consider a Lie algebra 
valued field\footnote{$\psi_1$ is a field in the complexified Lie
algebra such that $\exp{(i \psi_1)} \in \MG$, e.g.~for $\MG = SU(N)$ the
field $\psi_1$ takes values in the set of hermitian matrices.} 
$\psi_1$ in the adjoint
representation transforming homogeneously under gauge
transformations. In the following we will refer to such a field as
``Higgs field''. We fix the gauge by requiring that 
for every $x \in M$ with $M$ being the space-time manifold 
$\psi_1^V(x)$ is in some closed convex subset $\CF$ (to be specified 
below) of the Cartan subalgebra $\CH$, i.e.~we are
looking for a gauge transformation $V$ with 
\begin{equation}
\label{fund-domain-1}
\psi_1^V(x) = V(x)^{-1} \psi_1(x) V(x) = h(x) \quad \mbox{and}  
\quad h(x) \in \CF \subset \CH 
\quad \forall x \in M \, .
\end{equation}
Let us emphasize that it is not sufficient to require $\psi_1^V(x)$ 
to be an element of the Cartan subalgebra $\CH$. This would leave the
group of Weyl reflections $\MW$ unfixed, which is given by reflections 
in $\CH$ at planes through the origin perpendicular to a root. 
In fact if $w \in \MW$ then $V(x) w$ also rotates $\psi_1(x)$ into the
Cartan subalgebra but the transformed field $\psi_1^{(Vw)}(x)$ will in
general differ from $\psi_1^V(x)$ ($\psi_1^{(Vw)}(x)$ is the image of 
$\psi_1^V(x)$ under the Weyl reflection $w \in \MW$). Therefore, in
order to fix the gauge transformed image of $\psi_1(x)$ uniquely one has
to restrict $\psi_1(x) = h(x)$ to the so-called fundamental 
domain $\CF \subset \CH$ which is given by the coset $ \CH / \MW$, 
i.e.~the fundamental domain $\CF$ is obtained by identifying all vectors
of the Cartan subalgebra $\CH$ which are related by Weyl reflections 
$w \in \MW$. It is well known that the Cartan subalgebra decomposes 
into Weyl chambers related to each other by Weyl reflections and the
fundamental domain $\CF$ can be identified with a specific Weyl chamber,
which we choose as 
\begin{equation}
\label{fund-domain-2}
\CF = \left\{ \rho \, | \quad ( \rho , \alpha_{(i)} ) \geq 0 \quad 
\mbox{for all simple roots } \alpha_{(i)} \right\} \,. 
\end{equation} 
From equation (\ref{roots-weights-2}) we obtain that every 
$\rho \in \CF$ can be uniquely written as a linear combination of
the fundamental co-weights $\mu_{(i)}^\vee$ with real and positive 
coefficients
\begin{equation}
\label{fund-domain-3}
\rho \in \CF 
\Leftrightarrow 
\rho = \sum_{k=1}^r c_k \mu_{(k)}^\vee \quad \mbox{with } 
c_k = ( \rho , \alpha_{(k)} ) \geq 0 \, , \quad k = 1, \ldots , r \, .
\end{equation}  
From its definition (\ref{fund-domain-1}) it follows that 
the matrix $V(x)$ is defined only up to
right-multiplication with a matrix $g(x)$ commuting with $h(x)$
\begin{equation}
\label{residual}
V(x) \rightarrow V(x) g(x) \,  , \quad
g(x) h(x) g(x)^{-1} = h(x) \, .
\end{equation}
The set of all such matrices $g(x)$ form a subgroup of $\MG$, the 
centralizer
of $h(x)$ in $\MG$, denoted by $\MC_{h(x)} (\MG)$. The centralizer
contains the Cartan subgroup $\MH$ of $\MG$. At points $x$ where the
centralizer is just $\MH$ we can choose $V(x)$  and $h(x)$ smoothly, 
assuming the Higgs field $\psi_1(x)$ was smooth before gauge fixing. 
However, at points $x \in M$ for which $\psi_1^V(x) = h(x)$ is on the
boundary of the fundamental domain $\CF$ the residual gauge freedom is
enlarged, i.e.~the centralizer $\MC_{h(x)} (\MG)$ becomes non-Abelian 
and as a consequence there are obstructions to a smooth choice of $V(x)$
and $h(x)$. The set of such singular points (or gauge fixing defects)
\begin{equation}
\CD_\ag := \left\{ x \in M \, | \quad \MC_{h(x)} \neq \MH \right\} 
\end{equation}
is referred to as Abelian gauge fixing defect manifold. 
Generically the defect manifold
consists of connected subsets of co-dimension $3$, i.e.~they form
points in $D=3$ and lines in $D=4$ and represent magnetic monopoles and
monopole loops, respectively.

To illustrate that the centralizer of a point $\rho$ on the boundary 
of $\CF$ is non-Abelian, we  decompose $\rho$ as in equation
(\ref{fund-domain-3}). For $\rho$ on the boundary of $\CF$ at least one 
of the coefficients $c_k$ vanishes, say $c_l = 0$. Then it follows
\begin{eqnarray}
[ \rho , E_{\pm \alpha_{(l)}} ] &=& 
\sum_k \rho_k [ H_k , E_{\pm \alpha_{(l)}} ] = 
\sum_k \rho_k ( \pm (\alpha_{(l)})_k ) 
E_{\pm \alpha_{(l)}} 
\\
\label{su(2)-subgroup}
&=& 
\pm ( \rho , \alpha_{(l)} ) E_{\pm \alpha_{(l)}} = 
\pm c_l E_{\pm \alpha_{(l)}} = 0 
\end{eqnarray}
implying that the $SU(2)$-subgroup of $\MG$ generated by 
$E_{\pm \alpha_{(l)}}$ and $\alpha_{(l)}$ is contained in the
centralizer of $\rho$, i.e.~the centralizer is non-Abelian.
If there are more than one vanishing coefficients $c_k$ in
(\ref{fund-domain-3}), the centralizer becomes larger.  
A complete classification of the various types of possible centralizers
can be found in ref.~\cite{Ford-98-3}.

To identify magnetic monopoles and their charges we introduce the
magnetic gauge potential $A_{\mag}$ and its field strength $F_\mag$
\begin{equation}
\label{A-mag}
A_\mag := V^{-1} \d V _{| \CH} \, , \quad 
F_\mag = \d A_\mag = - V^{-1} \d V \wedge V^{-1} \d V _{| \CH} \, ,
\end{equation}
where $| \CH$ denotes projection onto the Cartan subalgebra $\CH$.
The gauge potential $A_\mag$ transforms as a gauge potential with
respect to the residual Abelian gauge transformations, see equation 
(\ref{residual}).
The magnetic charge of a defect is given by the flux through a closed
surface $\CS$ surrounding the defect\footnote{In $D=4$ the defect 
is generically a line. In this 
case one takes a $3$-dimensional space $\CK$ traversing the monopole line
in exactly one point, say $x_0$, and chooses $\CS$ to be a closed 
surface in $\CK$ surrounding $x_0$.} 
$\CS$ and integrate $F_\mag$ over $\CS$
\begin{equation}
Q_\mag = \frac{1}{2 \pi i} \int_\CS F_\mag
\end{equation}
The magnetic charge is quantized \cite{Ford-98-3}
\begin{equation}
\label{charge-quant}
Q_\mag = \sum_k n_k \alpha_{(k)}^\vee \, , 
\quad n_k \in {\mathbbm{Z}} \, , \quad k = 1 ,\ldots , r \, ,
\end{equation}
which is a generalization of the relation found in
refs.~\cite{Reinhardt-97-2,Jahn-98,Ford-98-1}.
Let us consider a defect at $x_0$ such that only one of the coefficients
$c_k$, say $c_l$, in the decomposition
\begin{equation}
\label{decomp-defect}
h(x_0) = \sum_{k=1}^r c_k \mu_{(k)}^\vee \, , 
\quad c_k > 0 \quad \mbox{for } k \neq l \, , 
\quad c_l = 0 
\end{equation}
vanishes. Then the charge $Q_\mag$ of the defect is an integer 
multiple of $\alpha_{(l)}^\vee$. In general one can show
that the coefficient $n_k$ in (\ref{charge-quant}) is zero if the
coefficient $c_k$ in (\ref{decomp-defect}) is non-zero.

\mysection{Center gauge fixing}

Above we have performed the Abelian gauge 
fixing in which the Higgs field $\psi_1$ has been gauge transformed into
the fundamental domain $\CF \subset \CH$
\begin{equation}
\psi_1^V (x) = h(x)  \in \CF \, .
\end{equation}
After this first step we are left with a residual Abelian 
$\MH \cong U(1)^r$ 
gauge freedom (away from defect points). This residual gauge freedom
will now be fixed up to the center of the gauge group. This is done
by rotating the non-Cartan part of the Higgs field $ \psi_2^V (x) $ in
the way described below:

We decompose the Higgs field $\psi_2^V(x) \in \CG$ 
with respect to a basis of the Lie algebra $\CG$: 
\begin{eqnarray}
\nonumber
\psi_2^V(x) 
&=& 
\sum_{k=1}^r h_k (x) \alpha_{(k)} 
\\
\nonumber
&& + 
\sum_{k=1}^r 
\left( e_{\alpha_{(k)}} (x) E_{\alpha_{(k)}}  + 
{e_{\alpha_{(k)}} (x)}^* E_{-\alpha_{(k)}} \right) 
\\
\label{decomp-1}
&& +
\sum_{\beta \in \Sigma^+ , |\beta| > 1} 
\left( f_\beta (x) E_{\beta} + 
{f_\beta (x)}^* E_{-\beta} \right) \, ,
\end{eqnarray}
where $e_{\alpha_{(k)}}(x)$ and 
$f_\beta(x)$ are complex numbers and the star
denotes complex conjugation. 
The first term on the r.h.s.~represents the part of $\psi_2^V(x)$ lying
in the Cartan subalgebra and the summation runs here over all simple
roots. Furthermore the second term represents the contributions from the
simple root vectors.
The center gauge fixing condition which fixes the residual Abelian 
(Cartan) gauge symmetries up to the center of the gauge group is chosen 
by requiring all coefficients of the simple roots, 
$e_{\alpha_{(k)}}(x) \, , \, k = 1 , \ldots , r$ to be real and positive. 

We will show that this requirement can be obtained by an Abelian
gauge transformation, leaving unfixed only the center of the gauge
group. Indeed an element $g(x) \in \MH$ of the Cartan subgroup 
can be parameterized by
\begin{equation}
\label{g-decomp}
g(x) = 
\exp \left( i \sum_k s_{k}(x) \mu^\vee_{(k)} \right) \, 
\end{equation}
where $\mu^\vee_{(k)}$ are the co-weights and the $s_k(x)$ are real
numbers.  

Adjoint action of $ g(x) $ on $ E_{\pm \alpha_{(k)}} $ yields 
by using equations (\ref{commutation-1}) and (\ref{roots-weights-2}),
i.e.
\begin{eqnarray}
[ \mu_{(l)}^{\vee} , E_{\alpha_{(k)}} ] &=& 
\left( \mu_{(l)}^{\vee} \right)_p [ H_p , E_{\alpha_{(k)}} ] = 
\left( \mu_{(l)}^{\vee} \right)_p 
\left( \alpha_{(k)} \right)_p E_{\alpha_{(k)}}
\\
&=& 
\left( \mu_{(l)}^{\vee} , \alpha_{(k)} \right) E_{\alpha_{(k)}} =
\delta_{k l} E_{\alpha_{(k)}} \, ,
\end{eqnarray}
the following relation
\begin{equation}
g(x)^{-1}  E_{ \pm \alpha_{(k)}} g(x) =  
\exp ( \mp i s_k(x) ) E_{\pm \alpha_{(k)}} \, .
\end{equation}
This means that the coefficients $e_{\alpha_{(k)}}(x)$ 
in equation (\ref{decomp-1})
transform under gauge transformation with $g(x)$ as
\begin{equation}
\label{e-trf}
e_{\alpha_{(k)}}(x) \rightarrow 
\exp ( - i s_k(x) ) e_{\alpha_{(k)}}(x) \, .
\end{equation}
If $e_{\alpha_{(k)}}(x) \neq 0$ the condition that 
$e_{\alpha_{(k)}}(x)$ to be real and 
positive after gauge rotation with $g(x)$ determines the 
parameters $s_k(x)$ modulo $2 \pi$. But 
$\exp{\left( 2 \pi i \mu^\vee_{(k)} \right)}$ is a 
center element (c.f.~equation (\ref{center-1})) of 
$\MG$. Therefore, requiring the coefficients of the simple roots, 
$e_{\alpha_{(k)}}(x)$, to be real and  positive fixes the group 
element $g(x)$ up to an element of the center of $\MG$.
This shows that the above proposed gauge condition indeed fixes the
gauge group $\MG$ up to its center.

\subsection{The center gauge fixing for specific gauge groups}

Below we illustrate the above proposed center gauge fixing for two
specific gauge groups.

\subsubsection{The gauge group $SU(3)$}

For gauge group $SU(3)$ this type of center gauge fixing is the one
suggested by de~Forcrand and Pepe \cite{deForcrand-00-1}. In this case 
the Higgs field $\psi_2^V(x)$ has the matrix representation
\begin{eqnarray}
\psi_2^V &=& 
\left(
\begin{array}{ccc}
h_1 & e_{\alpha_{(1)}} & f_{\alpha_{(1)} + \alpha_{(2)}} \\
e_{\alpha_{(1)}}^* & (-h_1 + h_2) & e_{\alpha_{(2)}} \\
f_{\alpha_{(1)} + \alpha_{(2)}}^* & e_{\alpha_{(2)}}^* & - h_2 
\end{array}     
\right)
\end{eqnarray}
and the
coefficients $e_{\alpha_{(k)}}(x)$ are simply the elements on the secondary
diagonal of the matrix $\psi_2^V(x) \in su(3)$. After center gauge
fixing these coefficients are demanded to 
be real and positive.
Gauge rotation with
\begin{eqnarray}
g &=& 
\exp \left( i 
( s_{1} \mu^\vee_{(1)} + s_{2} \mu^\vee_{(2)} ) 
\right) \\
&=& \diag 
\left( \exp(i/3(2 s_1 + s_2)), \exp(i/3(-s_1 + s_2)), 
\exp(i/3(-s_1 - 2 s_2)) \right)
\end{eqnarray}
transforms $\psi_2^V$ into
\begin{eqnarray}
g^{-1} \psi_2^V g &=&
\left(
\begin{array}{ccc}
h_1 & e_{\alpha_{(1)}} e^{-i s_1} & 
f_{\alpha_{(1)} + \alpha_{(2)}} e^{-i ( s_1 + s_2)}\\
e_{\alpha_{(1)}}^* e^{i s_1} & (-h_1 + h_2) & 
e_{\alpha_{(2)}} e^{-i s_2} \\
f_{\alpha_{(1)} + \alpha_{(2)}}^* e^{i(s_1+s_2)} & 
e_{\alpha_{(2)}}^* e^{i s_2} &  - h_2 \end{array} 
\right) \, .
\end{eqnarray}
Obviously, if $e_{\alpha_{(1)}} \neq 0$ and 
$e_{\alpha_{(2)}} \neq 0$ then $s_1$ and $s_2$ are defined 
modulo $2 \pi$. Since
\begin{eqnarray}
\exp \left( 2 \pi i \mu^\vee_{(1)} \right) 
&=& \diag \left( e^{4/3 \pi i} , e^{- 1/3 \pi i} , 
e^{-1/3 \pi i}  \right) =  e^{4/3 \pi i} \id \\
\exp \left( 2 \pi i \mu^\vee_{(2)} \right) 
&=& \diag \left( e^{1/3 \pi i} , e^{- 2/3 \pi i} , 
e^{-2/3 \pi i}  \right) = e^{1/3 \pi i} \id 
\end{eqnarray}
are just the center elements of $SU(3)$ 
this implies that $g$ is defined up to the 
center of $SU(3)$.
 

\subsubsection{The gauge group $Sp(2)$}

Below we will discuss the gauge group $Sp(2)$ which is the universal 
covering group of $SO(5)$ which is relevant in the context 
string theory, see e.g.~\cite{SO(5)}. The symplectic group $Sp(2)$ is
defined as the group of linear transformations in two-dimensional
quaternionic space leaving invariant the sesquilinear form 
\begin{equation}
({\mathbf a} , {\mathbf b} ) = \bar {\mathbf {a}}^t {\mathbf b} = 
\bar a_1 b_1 + \bar a_2 b_2 \in {\mathbbm{H}} \, ,
\end{equation}
where ${\mathbf a}$ is a quaternionic vector with the two quaternionic 
components $a_1$ and $a_2$. Here the bar denotes quaternionic 
conjugation and $t$ denotes transposition.
A group element $g$ of $Sp(2)$ is a $2 \times 2$ matrix with
quaternionic entries such that $\bar g^{t} g = \id$. A Lie algebra 
element $A$ of $sp(2)$ fulfills $\bar A^{t} + A = 0$.
If we represent quaternions by $2 \times 2$ matrices:
\begin{equation}
q = q_0 \id + i q_k \sigma_k \in {\mathbbm{H}} \, , \quad 
\bar q = q_0 \id - i q_k \sigma_k 
\end{equation}
the elements of $Sp(2)$ are $4 \times 4$ matrices 
made of $2 \times 2$ quaternionic blocks.
Then the decomposition (\ref{decomp-1}) of the Higgs field 
$\psi_2^V$ in terms of the roots reads
\begin{eqnarray}
\psi_2^V &=& 
\left(
\begin{array}{cccc}
( h_1 - \frac{1}{2} h_2 ) & e_{\alpha_{(1)}} 
& i e_{\alpha_{(2)}}^* \frac{1}{\sqrt{2}} 
& i f_{\alpha_{(1)}+\alpha_{(2)}} \frac{1}{\sqrt{2}} 
\\
e_{\alpha_{(1)}}^* & ( - h_1 + \frac{1}{2} h_2 ) &
- i f_{\alpha_{(1)}+\alpha_{(2)}}^* \frac{1}{\sqrt{2}} 
& i e_{\alpha_{(2)}} \frac{1}{\sqrt{2}} 
\\
- i e_{\alpha_{(2)}} \frac{1}{\sqrt{2}} 
& i f_{\alpha_{(1)}+\alpha_{(2)}} \frac{1}{\sqrt{2}} & 
\frac{1}{2} h_2 & f_{\alpha_{(1)}+2 \alpha_{(2)}} 
\\
- i f_{\alpha_{(1)}+\alpha_{(2)}}^* \frac{1}{\sqrt{2}} 
& - i e_{\alpha_{(2)}}^* \frac{1}{\sqrt{2}} & 
f_{\alpha_{(1)}+2 \alpha_{(2)}}^* & - \frac{1}{2} h_2
\end{array}
\right)
\end{eqnarray}
A short calculation shows that gauge rotation with 
\begin{eqnarray}
g &=& \exp{ 
\left( 
i ( s_{1} \mu^\vee_{(1)} + s_{2} \mu^\vee_{(2)} ) 
\right)} 
\\
\nonumber
&=& \diag 
\left( 
\exp(i/2 \, s_1), \exp(-i/2 \, s_1), 
\exp(i/2 \, ( s_1 + 2 s_2)) , \exp(- i/2 \, ( s_1 + 2 s_2)) 
\right)
\end{eqnarray}
yields
\begin{eqnarray}
e_{\alpha_{(1)}} &\rightarrow& 
e_{\alpha_{(1)}} e^{-i s_1} \\
e_{\alpha_{(2)}} &\rightarrow& 
e_{\alpha_{(2)}} e^{- i s_2} \, 
\end{eqnarray} 
as expected from equation (\ref{e-trf}).
Again $s_1$ and $s_2$ are fixed modulo $2 \pi$. 
Since furthermore
\begin{eqnarray}
\exp{( 2 \pi i \mu^\vee_{(1)})} &=& - \id \, , \\
\exp{( 2 \pi i \mu^\vee_{(2)})} &=& \id \, 
\end{eqnarray}
(where $\id$ denotes the $4$-dimensional unit matrix) represent the
center elements of $Sp(2)$ it follows that the center of $Sp(2)$ 
is unfixed by our gauge condition.


One further remark is here in order. 
One could also try to fix
the gauge by demanding $e_{\alpha_{(1)}}$ and 
$f_{\alpha_{(1)}+2 \alpha_{(2)}}$ to be real and 
positive. This corresponds to considering the $SU(2) \times SU(2)$ 
subgroup of $Sp(2)$ (given by diagonal quaternionic matrices) and 
center gauge fixing in each of the two $SU(2)$ subgroups. In this 
case the residual gauge freedom would be the product
of the centers of the two $SU(2)$ subgroups.
But this would imply that the residual gauge
freedom would be bigger than the center of $Sp(2)$ - the gauge
rotation $g$ would be fixed up to multiplication with one of the 
four group elements
\begin{equation}
\left(
\begin{array}{cc}
\id & 0 \\
0 & \id 
\end{array}
\right) \, , \quad
\left(
\begin{array}{cc}
-\id & 0 \\
0 & -\id 
\end{array}
\right) \, , \quad
\left(
\begin{array}{cc}
-\id & 0 \\
0 & \id 
\end{array}
\right) \, , \quad
\left(
\begin{array}{cc}
\id & 0 \\
0 & -\id 
\end{array}
\right) \, , 
\end{equation}
(where $\id$ denotes now the $2 \times 2$ unit matrix)
while the center of the group $Sp(2)$ is given by the first two matrices
only.

\mysection{Discussion of the gauge fixing defects}

In the course of the above described gauge fixing procedure 
different types of singularities can appear. 

Magnetic monopole defects arise whenever the residual gauge freedom 
after the first step of the gauge fixing procedure is larger than 
$\MH \cong U(1)^r$, 
i.e.~if the centralizer of $\psi_1^V(x)$ is larger than 
$U(1)^r$. This is equivalent to $\psi_1^V(x)$ lying on the boundary 
of the fundamental domain $\CF$. 

Vortex defects arise whenever the residual gauge freedom, which is left 
after the second step of the gauge fixing procedure, is larger 
than the center of
$\MG$. This is the case whenever one of the coefficients 
$e_{\alpha_{(k)}}(x)$ in the decomposition (\ref{decomp-1}) is 
zero\footnote{For gauge group $SU(2)$ this means that at such points 
$\psi_1(x)$ and $\psi_2(x)$ are linearly dependent.}.
The set of those points for which this happens is referred to in the
following as center gauge fixing defect manifold 
\begin{equation}
\CD_\cg := \left\{ x \in M \, | \quad 
e_{\alpha_{(k)}}(x) = 0 \quad \mbox{for at least one } k \, , 
1 \leq k \leq r \right\}
\end{equation}
since at these points the above described gauge fixing procedure is 
ill-defined. It is easy to see that connected subsets of this defect
manifold form vortices. Indeed, the equation 
$e_{\alpha_{(k)}}(x) = 0$ implies two conditions (one for the real and
one for the imaginary part of $e_{\alpha_{(k)}}(x)$). Therefore
vortex defects have co-dimension $2$, i.e.~vortex defects are
$1$-dimensional lines in $D=3$ and $2$-dimensional faces in $D=4$. 
If $e_{\alpha_{(k)}}(x)=0$ then the coefficient $s_k(x)$ in the 
decomposition (\ref{g-decomp}) is not fixed and the residual gauge
freedom is larger than the center of $\MG$ (the group of residual 
gauge transformations contains e.g.~the $U(1)$ subgroup generated by 
$\mu^\vee_{(k)}$).

Now we will show that the magnetic monopoles lie on top of vortices 
by construction. Let $\psi_1^V(x_0) = \rho$ be on the boundary 
of $\CF$ such that at least one of the coefficients in the 
decomposition (\ref{fund-domain-3}), say $c_l(x_0)$,  vanishes. 
Then the centralizer 
$\MC_{\rho} (\MG)$ contains the $SU(2)$-subgroup $\MK$ generated 
by $E_{\pm \alpha_{(l)}}$ and $\alpha_{(l)}$, c.f.~equation 
(\ref{su(2)-subgroup}).
Indeed we can choose a matrix $g \in \MK \subset \MC_{\rho} (\MG)$ 
such that in the decomposition (\ref{decomp-1}) of 
$\psi_2^{Vg}(x_0)$ the coefficient 
$e_{\alpha_{(l)}}(x_0)$ is zero, which means
that $x_0$ is also a defect of the center gauge fixing. In this sense we
obtain $\CD_\ag \subset \CD_\cg$. On the other hand 
from a physical point of view it is clear that from a magnetic monopole
with non-zero charge there must emanate a vortex (or Dirac string) 
carrying away the magnetic flux - and the vortex singularities 
introduced by our gauge fixing procedure indeed carry magnetic flux as we 
will show below. More rigorously one can argue as
follows: 

Let us assume there is a magnetic monopole with non-zero 
charge at $x_0$ and let us consider a surface $\CS$ 
surrounding $x_0$ such that the integral of $F_\mag$ over $\CS$ is
non-zero. Then it is impossible to choose the gauge transformation 
$Vg$ smoothly on the whole surface $\CS$. On the other hand let us 
assume that there is no vortex singularity traversing $\CS$. But then, 
because $\CS$ is simply connected, it is possible to choose the gauge 
transformation $Vg$ smoothly on $\CS$. But this contradicts the
assumption 
that the magnetic monopole at $x_0$ has non-zero charge. Hence, there
must be vortex singularities traversing $\CS$. As we can choose $\CS$
arbitrarily closed to $x_0$ we can conclude that a vortex singularity
emanates from the magnetic monopole at $x_0$.

In the following we will show that the flux of
$A_\mag$ (\ref{A-mag})
carried by a vortex is quantized - it is given by 
a linear combination of the co-weights with integer valued 
coefficients. For this purpose we will integrate $A_\mag$ 
along an infinitesimal loop
$\CC$ surrounding the vortex singularity\footnote{In $D=4$ the 
vortex is a two-dimensional surface. In this case one takes a 
$2$-dimensional face $\CK$ intersecting the vortex (singularity) 
sheet at exactly one point $x_0$ and chooses $\CC$ as a loop in $\CK$
surrounding $x_0$.} away from magnetic monopoles. We obtain
\begin{eqnarray}
\Phi &=& \frac{1}{2 \pi i} \int_\CC A_\mag = 
\frac{1}{2 \pi i} 
\int_\CC \left( (Vg)^{-1} \d (Vg)  \right)_{| \CH} \\
&=&
\frac{1}{2 \pi i} 
\int_\CC \left( g^{-1} V \d V g + g^{-1} \d g \right)_{| \CH} \, .
\end{eqnarray}
The integral over $V^{-1} \d V$ (first term
in the second line) vanishes, because we can choose $V$
smoothly on the whole path $\CC$, if the path is away from magnetic
monopoles. Hence we obtain
\begin{eqnarray}
\Phi &=& 
\frac{1}{2 \pi i} 
\int_\CC \left( g^{-1} \d g \right)_{| \CH} 
= 
\frac{1}{2 \pi} 
\int_\CC \left( \sum_{k=1}^r \d s_k \mu^\vee_{(k)} 
\right) \\
&=&
\frac{1}{2 \pi} 
\sum_{k=1}^r \Delta s_k \mu^\vee_{(k)} \, ,
\end{eqnarray} 
where $\Delta s_k$ is an integer multiple of $2\pi$. This is 
because $g$ at the
starting point of the loop $\CC$ can differ from $g$ at the endpoint 
only by a center element (\ref{center-1})
of $\MG$ ($s_k$ is fixed modulo $2 \pi$ by the
gauge fixing conditions).  

Integration of $F_\mag$ over an infinitesimal surface $\CS$ surrounding 
a magnetic monopole yields the continuity equation for magnetic flux.
The magnetic charge of the monopole is the sum of the fluxes of all
vortex singularities emanating from the monopole:
\begin{equation}
Q_\mag = \sum_l \Phi_l \, .
\end{equation}

The center gauge fixing can be visualized geometrically in a 
bundle picture. Appending to each 
$x \in \CD_\cg^c = M \setminus \CD_\cg$ all matrices $V(x) g(x)$, which
transform the Higgs fields $\psi_{1}$ and $\psi_2$ in the demanded way,
we get a principal bundle $P_\cg$ with structure group $\MZ(\MG)$, the
center of $\MG$, because $V(x) g(x)$ is fixed up to multiplication with
center elements. If the center of $\MG$ has $v$ elements, then $P_\cg$
is a $v$-fold covering of $\CD_\cg^c$. In analogy to complex function
theory we may look at $P_\cg$ as a Riemann surface of a multivalued
function. The vortex singularities can be identified as branching
points. Now we can classify the different vortex singularities. We
consider a closed loop $\CC$ surrounding the vortex singularity and lift 
this loop into the covering manifold $P_\cg$. There are $v$ different 
classes of lifted loops - the starting point and the endpoint of the 
lifted loops have to differ by one of the $v$ center elements, say $z$. 
If the center element $z$ is 
the identity (then the lifted loop is closed) we call the singularity a
Dirac string - otherwise we call it a center vortex.  
The center element $z$ is simply given by
\begin{equation}
z = \exp ( 2 \pi i \Phi ) \, ,
\end{equation}
where $\Phi$ is the magnetic flux through the vortex. 
On the other hand the center element $z$ is obviously equal to the 
Wilson loop\footnote{The $1$-form $\omega = (Vg)^{-1} \d (Vg)$ 
is invariant under changing $(Vg)$ by a center element $z$, 
i.e.~under the transformation 
$(Vg) \rightarrow (Vg) z \, , \, z \in \MZ(\MG)$. Therefore the form 
$\omega$ can be defined smoothly on the whole path $\CC$ around the 
center vortex.} 
\begin{equation}
\label{Wilson-loop}
z = \CP \exp{\left(
\int_\CC \omega (Vg)^{-1} \d (Vg)
\right)} \, \omega = (Vg)^{-1} \d (Vg) \, .
\end{equation}

We are interested in a globally well defined gauge transformation $Vg$
on $\CD_\cg^c$. But if we go once around a center vortex along a path
$\CC$ then the gauge transformation $Vg$ has to jump by a center element
at one point on the path $\CC$. This means that we have to introduce
cuts connecting the center vortices and on these cuts the gauge
transformation $Vg$ jumps by a center element. 

We can get rid of these cut singularities by working in the adjoint
representation, i.e.~by working with the gauge group $\MG / \MZ( \MG )$ 
instead of $\MG$. In this case the set $\{ g , g z_1 , g z_2 , \ldots \}$ 
of matrices in $\MG$ differing by
multiplication with center elements $ z_1,z_2,\ldots$ 
are identified as the element $\hat g$ in
the group $\MG / \MZ( \MG )$. But this means that the cuts where 
$Vg$ jumps by a center element in the fundamental representation 
are invisible in the adjoint representation $\hat V \hat g$. As a
consequence the gauge transformation $\hat V \hat g$ can be chosen 
smoothly everywhere except at center vortices, Dirac strings and 
magnetic monopoles. 
But the type of a center vortex singularity is no longer encoded 
in the Wilson loop\footnote{Strictly speaking one has to 
replace $\omega$ in equation (\ref{Wilson-loop}) by 
$\hat \omega = (\hat V \hat g)^{-1} \d (\hat V \hat g)$, i.e.~by 
$\omega$ in the adjoint representation.} (\ref{Wilson-loop})
along the path $\CC$ surrounding the vortex, since this Wilson loop 
always equals the identity in $\MG / \MZ( \MG )$. In the adjoint
representation the type of a center vortex can be related to an element 
of the first homotopy group $\pi_1(\MG / \MZ( \MG ))$ of the group 
$\MG / \MZ( \MG )$. To make this explicit we define a closed path 
$\hat \CC$ in the Lie group $\MG / \MZ( \MG )$ defined by
\begin{equation}
\label{Wilson-loop-adj}
\hat \CC (t) := 
\CP \exp{\left(
\int_0^t (\hat V(\CC(\tau)) \hat g(\CC(\tau)))^{-1} \,
\frac{\d (\hat V(\CC(\tau)) \hat g(\CC(\tau)))}{\d \tau} \, \d \tau 
\right)} \, ,
\end{equation}
where $\CC$ is a loop in the space time $M$ surrounding the center
vortex and it is parameterized by $t \, , 0 \leq t \leq 2 \pi$.
Obviously we obtain 
$\hat \CC(2 \pi) = \hat \CC(0) = \id_{\mathrm adj}$ which means
that the path $\hat \CC$ is closed in $\MG / \MZ( \MG )$. 
But the loop $\hat \CC$ is not contractible in $\MG / \MZ( \MG )$, if 
$\CC$ surrounds a center vortex. In the adjoint representation 
the vortex is classified by the homotopy class of the loop 
$\hat \CC$.

\section{Concluding remarks}

In the present paper we have given a prescription for center gauge
fixing for arbitrary gauge groups. The different types of singularities 
(magnetic monopoles and center vortices) appearing during the gauge 
fixing procedure have been discussed. Lattice calculations show that 
center vortices are
the relevant infrared degrees of freedom to explain color
confinement. In the continuum Yang-Mills theory the occurrence of center
vortices depends on topological properties of the gauge group - 
center   vortices are related (and classified) by elements of the center
of the gauge group or by elements of its first homotopy group. 
From this point of view it would be interesting to analyze a gauge group
which has a trivial center {\em and} a trivial homotopy group such as
the exceptional groups $G_2$ or $E_8$. Such groups should not show color
confinement, if center vortices are the only relevant infrared degrees 
of freedom for confinement.

\section*{Acknowledgments}

One of the authors (H.R.) is grateful to the Center of Theoretical
Physics, in particular to John Negele, for the hospitality extended to
him during his stay at MIT where this work was completed.

\end{document}